\documentclass[final,english]{bullsrsl}[2023/05/30]
\usepackage{amsmath}
\usepackage[latin1]{inputenc}
\usepackage[T1]{fontenc}

\usepackage{natbib} 
\usepackage{graphicx}
\usepackage{xcolor}
\usepackage{hyperref}

\hypersetup{
  colorlinks,
  citecolor=blue,
  linkcolor=blue,
  urlcolor=blue}

\begin{document}

\title{Study of blue metal-poor stars using UVIT/\textit{AstroSat}}

\author[affil={1}, corresponding]{Anju}{Panthi}
\author[affil={2}] {Annapurni}{Subramaniam}
\author[affil={1}]{Kaushar}{Vaidya}
\author[affil={3}]{Vikrant}{Jadhav}
\author[affil={2}]{Sharmila}{Rani}
\author[affil={2}]{Sivarani}{Thirupathi}
\author[affil={4}]{Sindhu}{Pandey}
\author[affil={5}]{Snehalata}{Sahu}

\affiliation[1]{Department of Physics, Birla Institue of Technology and Science, Pilani, Rajasthan-333031, India}
\affiliation[2]{Indian Institute of Astrophysics, Sarjapur Road, Koramangala, Bangalore, India}
\affiliation[3]{Helmholtz-Institut fur Strahlen-und Kernphysik, Universitat Bonn, Nussallee 14-16, D-53115 Bonn, Germany}
\affiliation[4]{Aryabhatta Research Institute of Observational Sciences, Manora Peak, Nainital, India}
\affiliation[5]{Department of Physics, University of Warwick, Coventry CV4 7AL, United Kingdom}

\correspondance{p20190413@pilani.bits-pilani.ac.in}
\maketitle

\begin{abstract}
Blue metal-poor stars are main-sequence stars that are bluer and brighter than typical turn-off stars in metal-poor globular clusters. They are thought to have either evolved through post-mass transfer mechanisms as field blue straggler stars or have accreted from Milky Way dwarf satellite galaxies. It has been found that a considerable fraction of blue metal poor stars are binaries, possibly with a compact companion. We observed 27 blue metal poor stars using UV imaging telescope of \textit{AstroSat} in two far-UV filters, F148W and F169M. In this work, we explain the possible formation channels of two stars, BMP17 and BMP37. We fit BMP17 with a single-component spectral energy distribution whereas BMP37 with a binary-component spectral energy distribution. As both of them are known SB1s, we suggest that the WD companion of BMP17 may have cooled down so that it is out of UV imaging telescope detection limit. On the other hand, we discover a normal mass white dwarf as the hot companion of BMP37, indicating mass transfer as the possible formation channel.
\end{abstract}


\section{Introduction}

The mysterious stellar populations known as blue straggler stars (BSS) were first discovered by \cite{sandage1953color} in the globular cluster (GC) M3. They are core-hydrogen burning stars that are brighter and bluer than the main sequence turn-off (MSTO) in the colour-magnitude diagram (CMD). The most plausible explanation of their formation includes mass transfer (MT) by Roche lobe overflow in primordial binary systems \citep{mccrea1964extended, 1976ApJ...209..734Z}, merger or MT of an inner binary in a triple system \citep{perets2009triple}, and direct stellar collisions \citep{hills1976stellar, chatterjee2013stellar}. Depending on the location of the primary star while transferring mass to the secondary, the MT mechanism can be further divided into three categories: Case A: when the primary star is on the main sequence (MS, \citealt{webbink1976evolution}), Case B: when the primary star is on the red giant branch phase (RGB, \citealt{mccrea1964extended}), and Case C: when the primary star is on the asymptotic giant branch phase (AGB, \citealt{chen2008binary}).

Different environments such as GCs \citep{sandage1953color}, open clusters (OCs, \citealt{ahumada1995catalogue}), the field population of the Milky Way \citep{preston2000these}, and dwarf galaxies \citep{momany2007blue} have all been found to contain BSS. Contrary to BSS identification in star clusters, field BSS (FBSS) identification is more challenging. \cite{preston1994space} discovered field metal-poor stars with MS gravities that were bluer than the MSTO of the GCs of equivalent metallicity, and were comparable to the BSS seen in clusters. Later, \cite{preston2000these} discovered that a major fraction of the stars of their sample to be binaries, and suggested that a significant portion of blue-metal poor (BMP) stars are FBSS created via MT.

It is well known that FBSS differ significantly from BSS found in star clusters in terms of surface chemistry \citep{andrievsky1996chemical, carney2005metal}. It is highly likely that the binary evolution in isolation and the progenitor properties will have a significant impact in the specifics of MT as well as the chemistry of the accreted material. Therefore, it is essential to find and characterise their companions in order to fully understand the formation of BMP stars and confirm whether they are true FBSS created via MT. The Ultraviolet Imaging Telescope (UVIT) onboard \textit{AstroSat} has played a remarkable role in the identification of hot companions of BSS in OCs and GCs \citep{subramaniam2016hot,sindhu2019uvit,sahu2019detection,singh2020peculiarities,jadhav2021uocs,vaidya2022uocs,panthi2022uocs,rani2023uocs,dattatrey2023globules}. Although, no such study has been conducted to characterize the hot companions of BMP stars. We present the study of identification and characterisation of the hot companions of FBSS and hence constraint their formation mechanism.

\section{Observations and data reduction}
The data used in this study are obtained using UVIT, one of the payloads onboard \textit{AstroSat} \citep{singh2014astrosat, agrawal2017astrosat}, India's first multi-wavelength space observatory, as well as other archival data. Readers are directed to \cite{kumar2012ultraviolet}, \cite{subramaniam2016orbit}, and \cite{tandon2017orbit} for additional information on the instrument and calibration results. We observed 27 BMP stars by UVIT using the F148W and F169M FUV filters with the exposure time of $\sim$ 200 sec to $\sim$ 2000 sec. Our choice of filters is guided by the reason that the excess due to an unresolved hot companion is prominent at wavelengths shorter than 1800 \AA, whereas the exposure times are proposed keeping in mind the brightness of the objects to obtain sufficient signal to noise ratio. In order to obtain the science-ready images, we employed CCDLAB \citep{postma2017ccdlab, postma2021uvit}. CCDLAB is a customized software which applies spacecraft drift corrections, geometric distortions corrections, flat field corrections, and astrometric corrections to the level 1 data and get the science ready images. We performed the aperture photometry on all the science-ready image using CCDLAB and followed curve-of-growth technique. We obtain the counts per pixel of the sources in both the FUV filters from the photometry and divided them by exposure times to obtain counts per second. These counts per seconds were then converted into magnitudes and fluxes using the zero point (ZP) magnitudes and unit conversions (UC) taken from the \cite{tandon2020additional}. Then we performed the aperture and saturation corrections following \cite{tandon2020additional}. The coordiantes and the observational details of the 2 BMP stars presented in this work are listed in Table \ref{Table1}.

\begin{table*}
\addtolength{\tabcolsep}{-3pt}
\caption{\footnotesize For BMP 17 and BMP 37, the coordinates in Columns 2-3, exposure times FUV filters in Coulmns 4-5, zero point magnitudes in Column 6, unit conversions in Coulmn 7, photometric magnitudes in F148W filter in Column 8, and photometric magnitudes in F169M filter in Column 9. }
\small
\begin{tabular}{ccccccccccc}
\hline
\hline
Name&RA&DEC&Exp. time &Exp. time&ZP&UC&F148W&F169M\\
\hline
~&(deg)&(deg)&(F148W)&(F169M)& & (x10$^{-15}$)&(mag)&(mag)\\
\hline 
BMP17&0.26576&$-$33.80442&1193.218&680.898&18.097$\pm$0.01&3.09$\pm$0.02&21.78$\pm$1.02&21.17$\pm$0.76\\
BMP37&8.96233&$-$17.95005&1183.078&966.068&17.410$\pm$0.01&4.392$\pm$0.03&19.42$\pm$0.73&18.89$\pm$0.14\\
\hline
\label{Table1}
\end{tabular}
\end{table*}

\section{Data analysis and results}
In order to construct the SEDs of BMP stars, we use virtual observatory SED analyzer (VOSA, \citealt{bayo2008vosa}). We obtain the photometric fluxes of all the stars in FUV and NUV wavelengths from GALEX \citep{martin2005galaxy}, optical from \textit{Gaia} DR3 \citep{brown2021gaia} and PAN-STARRS \citep{chambers2016pan}, near-infrared from Two Micron All-Sky Survey (2MASS, \citealt{cohen2003spectral}), and far-infrared from Wide-field Infrared Survey Explorer (WISE, \citealt {wright2010wide}) using VOSA. We provide the values of extinctions from \textit{Galactic Dust Reddening and Extinction map} \footnote{https://irsa.ipac.caltech.edu/applications/DUST/}.  
VOSA performs the extinction corrections in the observed fluxes using \cite{fitzpatrick1999correcting} and \cite{indebetouw2005wavelength} in all the wavelength bands based on the extinction values we provide. Furthermore, it selects the best possible SED fit by performing the $\chi^{2}$ test, by comparing the observed flux to the synthetic flux. For further details on the steps performed for SED fitting, readers are referred to \cite{panthi2022uocs}.

\begin{figure*}
\centering
\includegraphics[scale=0.25]{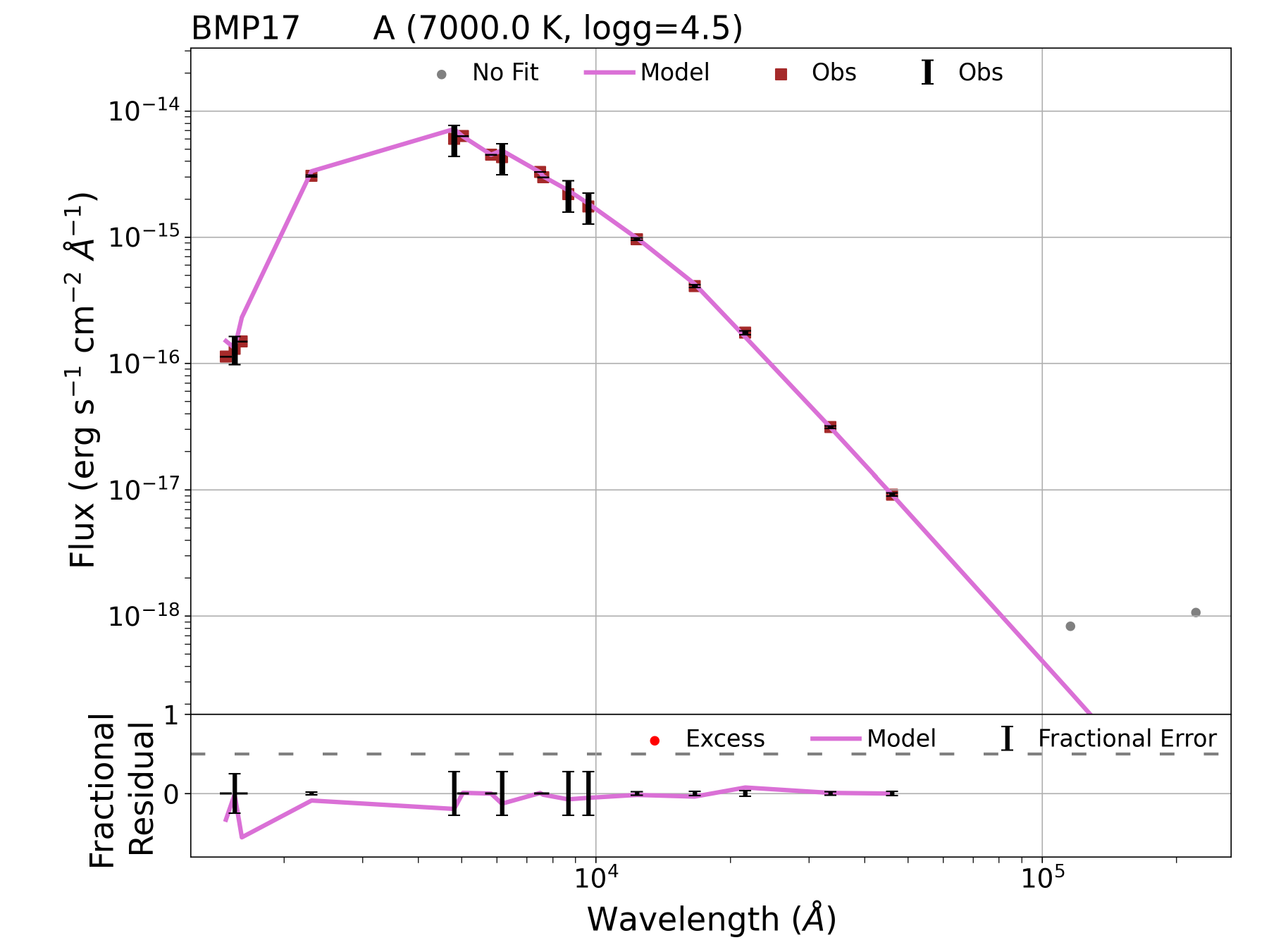} 
\includegraphics[scale=0.15]{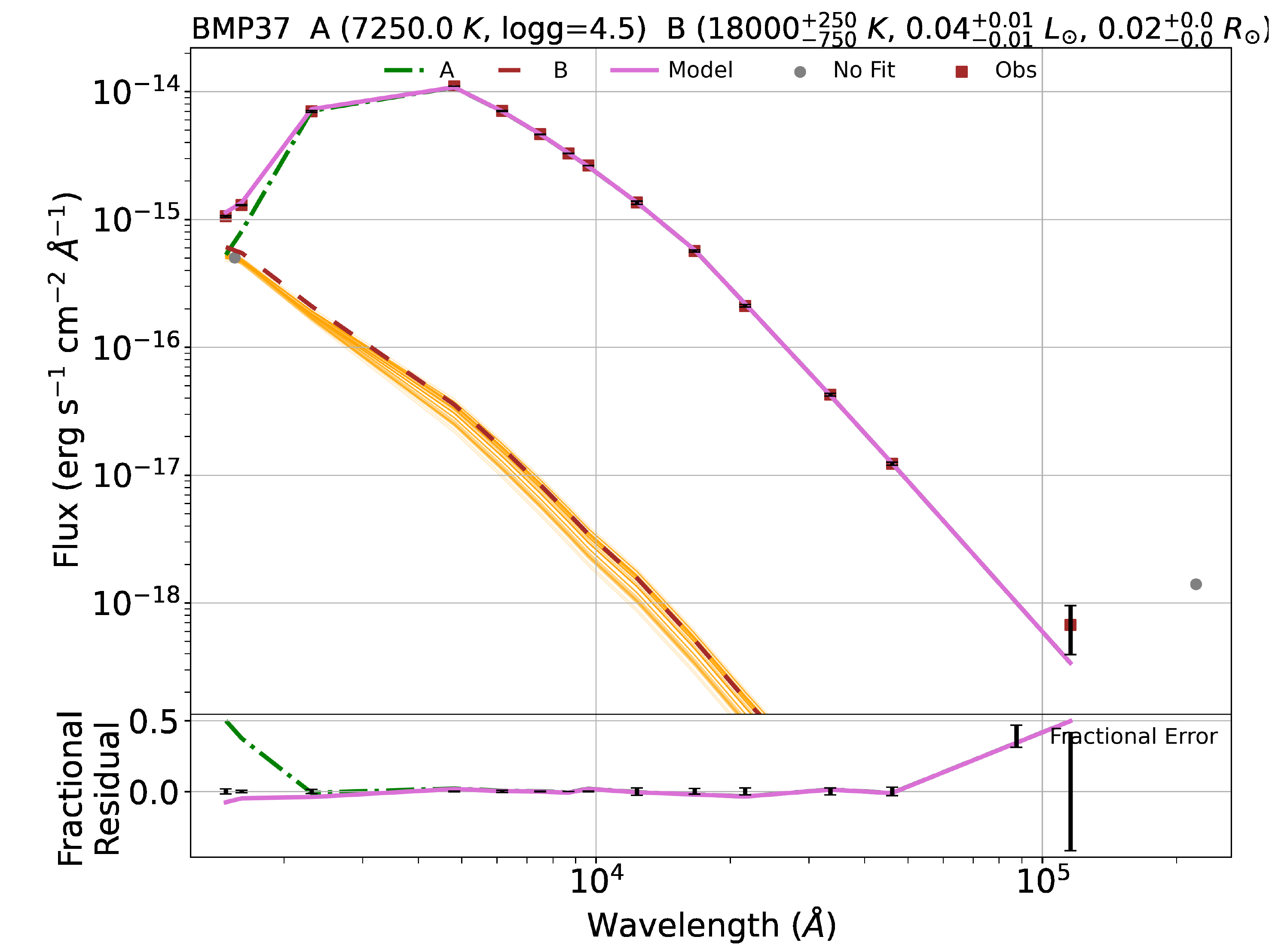}
\caption{\footnotesize SEDs of BMP17 and BMP37. In the top panel of single-component SED, brown data points represent the extinction corrected observed fluxes, with black error bars as the errors in observed fluxes, and the pink curve representing the Kurucz stellar model fit. The bottom panel shows the residual between extinction-corrected observed fluxes and the model fluxes across the filters from UV to IR wavelengths. In the binary-component SED, the top panel shows the model SED with the cooler (A) component in the green dashed line, hot (B) component in the brown dashed line along with the iterations shown as orange lines, and the composite fit in the pink solid line. The extinction corrected flux data points are shown as brown filled squares with error bars according to flux errors. The data points not included in the fit are shown as grey filled circles. The bottom panel shows the fractional residual for both single fit (green) and composite fit (pink). The parameters of BMP stars are mentioned at the top of the figure.}
\label{Fig.1}
\end{figure*}

We noticed that 10 BMP stars showed fractional residuals that were close to zero in all filters, thus we fitted single-component SEDs to those stars. On the other hand, we attempted to fit the double-component SEDs to 17 BMP stars after noticing an excess of more than 50$\%$ in their UV emissions. Among the 17 BMP stars that exhibited an excess in UV fluxes, we were able to accurately fit the binary component SEDs of 13 of them.
The SEDs of all the 27 BMP stars and their possible formation mechanisms are described in Panthi et al. (2023) (under review). Here we explain the formation mechanisms of two BMP stars, BMP17 and BMP37. In Figure \ref{Fig.1}, the left panel shows the SED of BMP17 which is fitted with a single-component SED, whereas the right panel shows the SED of BMP37, which is fitted with a binary-component. The parameters of the cool component are taken from VOSA and that of the hot-companion is determined using the python-code mentioned above. In case of hot companions, the errors were estimated by generating 100 iterations of observed SEDs by adding Gaussian noise proportional to the errors. We take the median of the parameters derived from the 100 SEDs as the parameters of hot companions, whereas we consider the standard deviation of the median as the error in the parameters.

\section{Discussion}

\begin{figure*}
\centering
\includegraphics[scale=0.38]{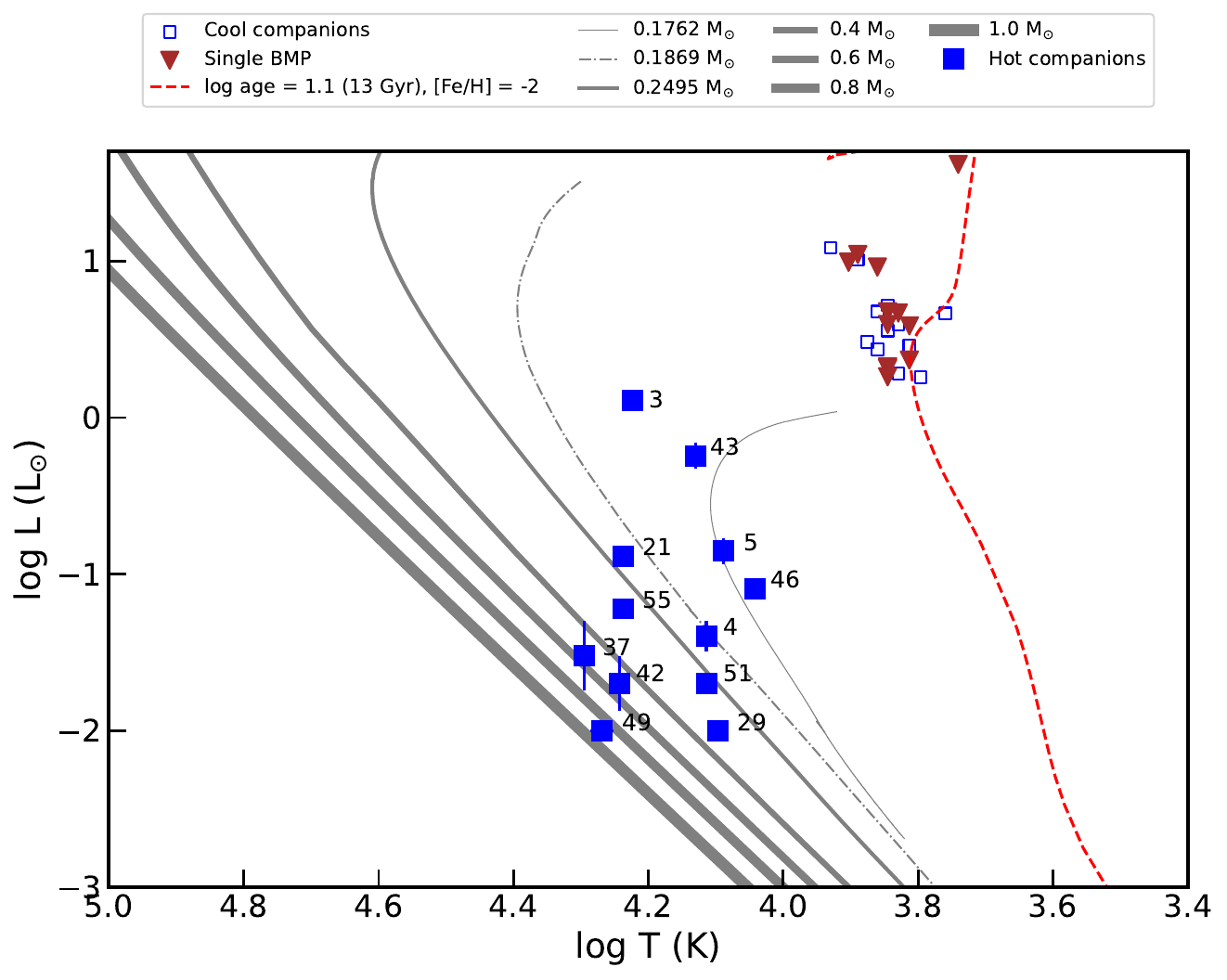}
\caption{\footnotesize The H-R diagram showing the single component BMP stars as brown triangles, the cooler companion of BMP stars as blue open squares, and their corresponding hot companions as blue filled squares. A PARSEC isochrone of 13 Gyr age is plotted as the red dashed curve. The WD cooling curves of different masses taken from \cite{panei2007full} and \cite{althaus2013new} are represented by different curves.}
\label{Fig.2}
\end{figure*}

The characteristics of the hot companions of BMP stars should be investigated in order to comprehend the formation processes of these stars. Therefore, we plotted the Hertzprung-Russell (H-R) diagram as shown in Figure \ref{Fig.2}. As can be seen, the hot companions of BMP stars are white dwarfs of extremely low-mass to normal mass. The details regarding the companions and formation mechanisms of the associated BMP stars is presented in Panthi et al. (2023). Here, we have presented a discussion on two BMP stars.

\textit{BMP 17 (BPS CS 22876-0021)} is a known SB1 with a period = 176.9 days and eccentricity = 0.1 \citep{preston2000these}. Due to the enhancement in \big[Sr/Fe\big] and \big[Ba/Fe\big], \cite{sneden2003binary} classified it as a FBSS. Since, we did not detect any UV excess, we suggest that this star may harbour a cooler WD. This implies that the MT must have happened long-ago and the cool WD is beyond the detection limit of UVIT. \\
\textit{BMP37 (GD 625)} has been classified as a SB1 with period = 84 days and eccentricity = 0.07 \citep{preston2000these}. Furthermore, it has been classified as FBSS, based on the enhancements in [Sr/Fe] and [Ba/Fe] \citep{sneden2003binary}. The best-fit SED parameters suggests that the hot companion to this star is a normal mass WD of mass $\sim$ 0.6M$_{\odot}$. We infer that this is a FBSS formed via Case-C MT, where the observed enhancements in \big[Sr/Fe\big] and \big[Ba/Fe\big] are in agreement with this discovery.

\section{Summary and conclusions}

\begin{enumerate}

\item We study 27 potential binary BMP stars using two FUV filters, F148W and F169M, of \textit{AstroSat}/UVIT. We construct their multi-wavelength SEDs and note that 10 BMP stars showed an excess $<$ 50$\%$, where 17 of them showed an excess $>$ 50$\%$ in UV fluxes. The detailed results of the 27 BMP stars are presented in Panthi et al. 2023 (under review).  
\item Out of the two BMP stars explained in this work, BMP17 is fitted with the single-component SED whereas BMP37 is fitted with the binary-component SED.
\item BMP17 is a known SB1 (period = 176.9 days and e = 0.1) and enhanced in \big[Sr/Fe\big] and \big[Ba/Fe\big]. We suggest that the absence of excess in UV fluxes is due to the fact that the MT is not recent. Hence, the WD has cooled down and therefore not detected using UVIT. 
\item BMP37 is also a known SB1 with a period of 84 days and e = 0.07. We discovered a normal-mass WD as the hot companion which indicates the Case-C MT. This finding is in agreement with the observed enhancements in \big[Sr/Fe\big] and \big[Ba/Fe\big].

\end{enumerate}

\textbf{Acknowledgements}\\

This publication uses the data from the AstroSat mission of the Indian Space Research Organisation (ISRO), archived at the Indian Space Science Data Centre (ISSDC).

\begin{furtherinformation}

\begin{authorcontributions}
AP$-$ Investigation, Methodology, Data curation, Formal Analysis, Visualization, Writing-original draft \\
AS$-$ Conceptualization, Supervision, Methodology, Validation, Investigation, Writing-review and editing \\
KV$-$ Supervision, Methodology, Validation, Investigation, Writing-review and editing \\
VJ$-$ Data curation, Software, Methodology, Visualization, Writing-review and editing \\
SR$-$ Data curation, Visualization, Writing-review and editing\\
ST$-$ Writing-review and editing \\
SP$-$ Data curation, Writing-review and editing\\
SS$-$ Data curation \\

\end{authorcontributions}

\begin{conflictsofinterest}
The authors declare no conflict of interest.
\end{conflictsofinterest}

\end{furtherinformation}
\bibliographystyle{bullsrsl-en}

\bibliography{extra}

\end{document}